# OFDM-Autoencoder for End-to-End Learning of Communications Systems

Alexander Felix*, Sebastian Cammerer*, Sebastian Dörner*, Jakob Hoydis†, and Stephan ten Brink*
* Institute of Telecommunications, Pfaffenwaldring 47, University of Stuttgart, 70659 Stuttgart, Germany
{felix,cammerer,doerner,tenbrink}@inue.uni-stuttgart.de
†Nokia Bell Labs, Route de Villejust, 91620 Nozay, France, jakob.hoydis@nokia-bell-labs.com

*Abstract*—We extend the idea of end-to-end learning of communications systems through deep neural network (NN)-based autoencoders to orthogonal frequency division multiplexing (OFDM) with cyclic prefix (CP). Our implementation has the same benefits as a conventional OFDM system, namely single-tap equalization and robustness against sampling synchronization errors, which turned out to be one of the major challenges in previous single-carrier implementations. This enables reliable communication over multipath channels and makes the communication scheme suitable for commodity hardware with imprecise oscillators. We show that the proposed scheme can be realized with state-of-the-art deep learning software libraries as transmitter and receiver solely consist of differentiable layers required for gradient-based training. We compare the performance of the autoencoder-based system against that of a state-of-the-art OFDM baseline over frequency-selective fading channels. Finally, the impact of a non-linear amplifier is investigated and we show that the autoencoder inherently learns how to deal with such hardware impairments.

## I. Introduction

Recently, end-to-end learning of communications systems has been proposed [1] based on deep neural networks (NNs), and in particular on the autoencoder concept ([2, Ch. 14]). In contrast to conventional communications systems, such a setup allows joint optimization of transmitter and receiver for any differentiable channel model without being limited to component-wise optimization. This approach breaks up restrictions commonplace in conventional block-based signal processing by moving away from handcrafted, carefully optimized sub-blocks towards adaptive and flexible (artificial) NNs, leading to many attractive research questions.

*"Classical"* block-based signal processing has shown to be close to optimal while each sub-block can be optimized individually for a specific task such as equalization, modulation, or channel coding [3]. At first glance, machine learning techniques do not appear to be a good match to communications on the physical layer, with 50 years of tremendous progress based on classic signal processing, communication and information theory, approaching close-to-optimal Shannon limit performance on many channels. However, several open problems remain, e.g., pertaining adaptivity and complexity of

We would like to thank Maximilian Arnold for many helpful discussions. We also acknowledge support from NVIDIA through their academic program which granted us a TITAN X (Pascal) graphics card.

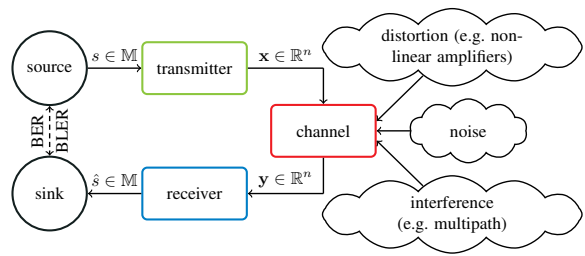

Fig. 1: Illustration of a simple communications system.

joint processing, where first results using machine learning-based approaches are promising (see [1], [4] and references therein). On the other hand, deep learning techniques have been shown to be very promising in scenarios where the channel is either (partially) unknown, or too complex for an analytical description [5], [6]. Furthermore, the benefits of machine learning approaches may include more flexible hardware, highly adaptive systems, and less overall complexity.

In our previous work [7], we demonstrated that an autoencoder-based system, solely composed of NNs, can communicate over-the-air without the need of any conventional signal processing block. However, an implementation without any conventional synchronization stage requires a lot of effort regarding synchronization for a single-carrier modulation scheme (see Section III in [7]).

The main contribution of this work is to extend our previous work to an orthogonal frequency division multiplexing (OFDM) scheme to enable reliable transmission over multi-path channels and to overcome the aforementioned synchronization problems. This allows learning of transmitter and receiver implementations—without any prior knowledge—that are optimized for an arbitrary differentiable end-to-end performance metric, e.g., block error rate (BLER). Besides more robustness against sampling synchronization errors, the OFDM system also benefits from the well-known feature of single-tap equalization over multipath channels.

From an implementation perspective, end-to-end training requires the availability of the gradient of each layer in the computation graph in order to apply gradient descent training. However, state-of-the-art deep-learning software libraries have already implemented a wide range of operations and its corresponding gradient. As a result, we can apply intermediate computations (without being trainable itself) on the signal such as the fast Fourier transform (FFT) without hindering the end-to-end training capabilities of the computation graph.

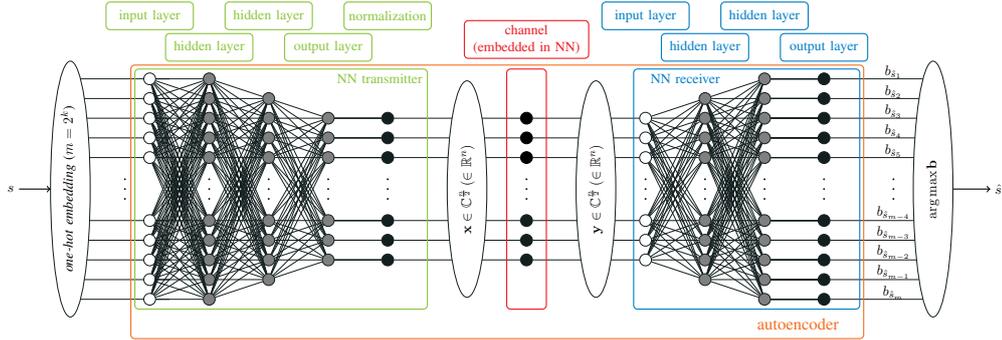

Fig. 2: Illustration of an end-to-end communications system as an autoencoder.

At first, it may appear counterintuitive to extend the flexible and general structure of the autoencoder by a fixed OFDM scheme. In other words, if this structure is helpful, the NN should be able to learn (its approximation) anyway. However, similar to previous work [1], we observe that injecting expert domain knowledge in the form of know transformations speeds-up training and increases performance. This is similar to the use of specialized network structures, such as convolutional neural networks (CNNs) in the field of computer vision [2], which despite reducing the expressive power of a NN leads to faster training, reduced complexity, and better generalization. Overall, the benefits of the proposed OFDM autoencoder are:

1) Robustness against sampling synchronization errors
2) Single-tap equalization
3) Moderate training complexity due to independent and short length sub-carrier messages
4) Compatibility with existing schemes
5) Robustness against hardware impairments, such as non-linear power amplifiers

## II. END-TO-END LEARNING OF A COMMUNICATIONS LINK

Next, we provide a short introduction to autoencoder-based communication. For further details, we refer the interested reader to [1], [7]. In an abstract view (see Fig. 1), a communications system consists of a transmitter, which seeks to transmit one message $s$ (out of $m$ possible messages) over a noisy channel to a receiver, whose task is to provide an estimate $\hat{s}$ of the noisy message. Assuming the system is allowed to have $n$ channel uses for the transmission of a single message, the resulting communication rate is $R = \log_2(m)/n$ in bits/channel use. As depicted in Fig. 2, one can implement both transmitter and receiver as deep NNs, separated by a channel (e.g., a multipath channel). In the machine learning context, this can be interpreted as an autoencoder [2], i.e., a NN which is trained such that its output $\hat{s}$ equals its input $s$. The intermediate noise layer requires the NN to find a robust representation of the information suitable for transmission over the actual channel. Note that the channel is also represented by network layers (operations, without trainable weights) that carry out stochastic transformations of the input data. It is crucial to have a model that accurately reflects the real channel as good as possible, however, as analyzed in [7], a mismatch can be tolerated through *finetuning* of the receiver. In an end-to-end stochastic gradient descent (SGD) training procedure, the receiver directly learns to recover the original transmitted information. After training, transmitter and receiver (i.e., their trained weights) can be deployed on the final application, e.g., on a software-defined radio (SDR) platform (see [7]).

As commonly done in the communications literature, and with slight abuse of notation, we interpret two consecutive real numbers at the output of the transmitter as a single complex number and revert this step at the receiver input. However, the channel could be also implemented with real-valued operations. Throughout this work, we use dense layers for transmitter and receiver, however, the receiver structure equals the *sequence detector* in [7], i.e., it uses a radio transformer network (RTN) as proposed in [8] to simplify the equalization procedure.

During training, the encoder part of the autoencoder has learned robust symbol sequence representations of all messages. Fig. 6 shows constellation diagrams of the IQ-symbols of all of the $m = 256$ possible messages of the single-carrier system, i.e., per subcarrier of the multi-carrier system. Each diagram shows all symbols at the same symbol position within a message, as each message consists of 4 complex-valued IQ-symbols (we assume $n = 8$ and consider half of the transmitter output as the real and the other half as the imaginary part).

### A. Synchronization Challenges

One of the major challenges we faced in our prior work [7], is that of handling the synchronization between transmitter and receiver for the autoencoder-based system caused by unsynchronized oscillators. This results in sampling frequency offset (SFO) and carrier frequency offset (CFO) making the receiver task demanding. Since the autoencoder system is operating directly on a measured IQ-sample stream, it additionally has to figure out the offsets of where a message starts and where it ends. Moreover, these offsets are not constant due to a dynamically changing SFO and we finally arrived at using an additional offset estimation NN, which is able to detect the beginning of a message within an IQ-sample stream with sufficient accuracy. An OFDM scheme in combination with a cyclic prefix (CP) solves these synchronization issues and simply provides (almost) perfectly sampled IQ-symbols to the receiver part of the autoencoder.

## III. OFDM EXTENSIONS

We extend our work of [7] from single-carrier to multi-carrier transmission, i.e., OFDM with CP as shown in Fig. 3. Note that a single autoencoder message **x** is represented by $\frac{n}{2}$ complex-valued IQ-symbols. Instead of directly transmitting the encoder's output **x**, an inverse discrete Fourier-transform (DFT) of width $w_{\text{FFT}}$ is applied on a set of $w_{\text{FFT}}$ independent autoencoder messages, i.e., $w_{\text{FFT}}$ equivalent independent sub-channels are created, where independent autoencoder messages are assigned to each subcarrier.[1] As each autoencoder still requires $\frac{n}{2}$ channel uses, we generate $\frac{n}{2}$ complex-valued OFDM symbols $\mathbf{x}_{\text{OFDM}}$, each of length $w_{\text{FFT}}$. For additional robustness against sampling synchronization errors and to avoid inter-symbol interference (ISI), we further add a CP of length $\ell_{\text{CP}}$, i.e., $w_{\text{FFT}}$ independent (in discrete *f-domain*) autoencoder symbols form one single OFDM symbol (in discrete *t-domain*) $\mathbf{x}_{\text{OFDM,CP}}$ of total length $w_{\text{FFT}} + \ell_{\text{CP}}$. Thus, a sequence of $\frac{n}{2}(w_{\text{FFT}} + \ell_{\text{CP}})$ complex-valued symbols is subsequently transmitted over the (multipath) channel.

At the receiver side, the CP can be used for frame synchronization through autocorrelation with peak detection; synchronization turned out to be a challenging step in single-carrier autoencoder-based communication [7] as already mentioned in the previous section. Finally, a DFT recovers the inputs for the $w_{\text{FFT}}$ independent autoencoder receivers.

The RTN equalizer [8], operates on a per-subcarrier basis on sequences of $n/2 + 1$ symbols, i.e., a message plus one fixed pilot tone. In some experiments, no explicit pilot is used, i.e., the decision is based on only $n/2$ symbols.

### A. Channel model

For our simulations, we choose a WiFi-inspired stochastic channel model. This model is derived from a Proakis *type C* tap-delay-line model with five coefficients [9]. For a fair comparison, it is important to challenge the autoencoder with changing channel conditions. Otherwise the NN would just learn the channel coefficients and optimize for a fixed system [10]. Therefore, we choose to use the Proakis coefficients as variances of normal-distributed channel taps. Further attributes of our channel are:
- AWGN
- Multipath-fading
- Only for Section IV-D: constant phase offset $\phi_{off}$ and incrementally increasing CFO over each symbol
- Only for Section IV-E: non linear amplifier with AM-AM distortion

In comparison with our previous work, we consider multipath propagation with five taps and do not have to consider sample time offset or synchronization due to our OFDM system with CP. Moreover, no explicit pulse shaping is needed as the FFT implies rectangular pulse shaping per sub-carrier. However, CFO must be considered, as the mismatch between the oscillator frequency of transmitter and receiver causes a phase offset, which in turn causes a rotation between the complex IQ samples over time.

[1]Remark: as no additional piloting is assumed, we cannot simply distribute the $\frac{n}{2}$ symbols of a message within the same OFDM symbol. Otherwise the unknown phase rotation per subcarrier would destroy the message.

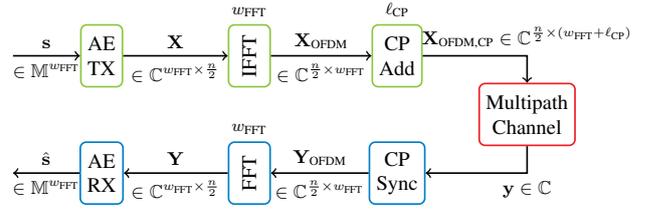

Fig. 3: OFDM extension to the autoencoder system.

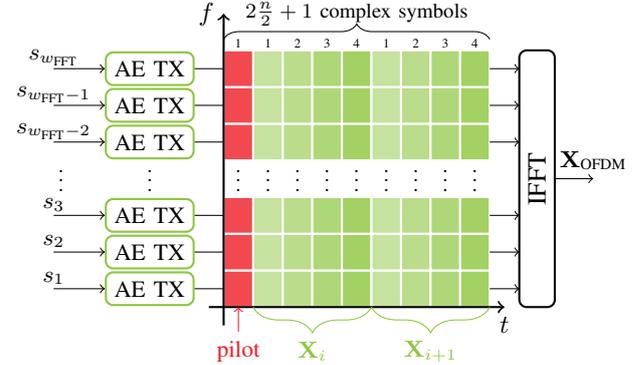

Fig. 4: OFDM Transmitter with pilot symbol every 2 messages in frequency (FFT width $w_{\text{FFT}}$) over time.

### B. Baseline

For the baseline system, we use minimum mean squared error (MMSE) channel equalization per sub-carrier (as it does not need channel and noise covariance matrices over sub-carriers) based on one pilot symbol $p \in \mathbb{C}$ per $2 \cdot \frac{n}{2}$ symbols of two messages $s$ (see Fig. 4), i.e., for the received symbol $y_i \in \mathbb{C}$ (in discrete *f-domain*) of sub-carrier $i$, the channel estimate is given as

$$\hat{h}_i = \frac{y_i \cdot p^*}{|p|^2 + \sigma^2}.$$

To provide a fair comparison, we compute the block error rate of our quadrature phase-shift keying (QPSK) baseline system on a block of $n/2$ symbols. Additionally, for CFO affected channels, we use a compensation based on the phase shift that can obtained from the CP.

## IV. SIMULATION RESULTS AND INTERPRETATION

Analyzing an autoencoder-based system in general is a non-trivial task as multiple effects come together and no distinct blocks can be identified. However, we conduct a few experiments to provide some insights on what specifically the system learns and how it behaves.

As shown in Fig. 4, we consecutively transmit two times $\frac{n}{2} = 4$ (i.e., $n = 8$) complex valued symbols and, when indicated, we add an additional pilot at the beginning. Further sys-tem parameters can be found in Tab. I. Training is performed in an end-to-end fashion without exhaustive hyperparameter optimization but manually optimized for suitable performance (see Tab I). Training the whole system from the scratch takes about half a day performed on a single NVIDIA TITAN X (Pascal) graphics card.

TABLE I: Parameters used for the autoencoder setup.

| Parameter | Value |
|---|---|
| Optimizer | SGD with Adam |
| Learning rate $L_r$ | 0.001 |
| Training SNR $E_b/N_0$ | 33dB and 23dB |
| Size of Dense Layers | 256 neurons |
| Number of Dense Layers: | |
| - Encoder: | 1 |
| - Decoder: | 2 |
| - Equalization: | 3 |
| Number of subcarriers $w_{\text{FFT}}$ | 64 |
| Size cyclic prefix $\ell_{\text{CP}}$ | 8 |
| Number of complex channel uses per message $n$ | 4 |
| Number of information bits per message $k$ | 8 |
| Number of messages $M$ | 256 |
| CFO Parameter: | |
| Carrier frequency $f_c$ | 2.35 GHz |
| Sample frequency $f_c$ | 20 MHz |
| Oscillator accuracy | 20 ppm |
| CFO maximum | 1.69°/sample |

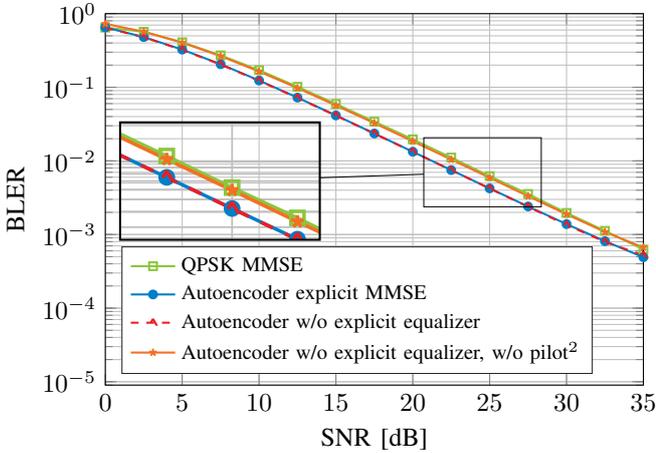

Fig. 5: BLER comparison of the autoencoder and QPSK baseline with MMSE equalization.[2]

### A. BLER-Results

As shown in Fig. 5, the autoencoder-based setup outperforms the QPSK baseline by approximately 2 dB over the whole SNR range, while having the same spectral efficiency. Note that the baseline could be extended to higher order modulation formats in combination with channel coding for further improvements. However, for these very short messages of only $k = 8$ information bits per block we do not expect large coding gains.

### B. Equalizing

Our goal is to give the NN most degrees of freedom and provide least explicit components as possible. Thus, we compare how well the single-tap equalizer can be emulated by the NN.

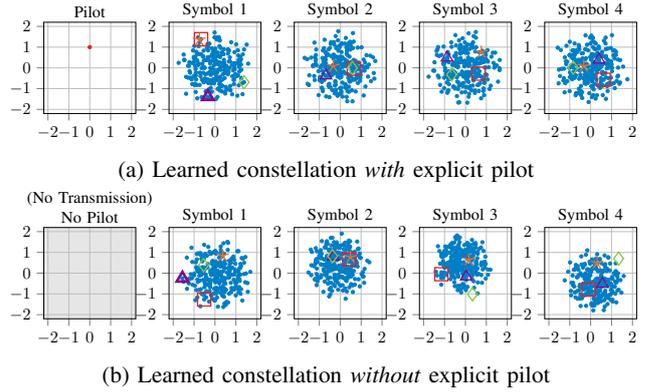

(a) Learned constellation *with* explicit pilot

(b) Learned constellation *without* explicit pilot

Fig. 6: Scatter plot of the learned constellations for all $m = 256$ messages using average power normalization $\|\mathbf{x}\|^2 \leq n$. The symbols of four individual messages are highlighted by different color markers.

Interestingly, no additional loss function is introduced within this block and further experiments showed no practical benefits in this attempt. We analyze the three different scenarios below and show the results in Fig. 5:

1) *Explicit pilot and explicit MMSE equalizer:* As a first proof of concept, we combine the autoencoder with a conventional MMSE equalizer. Our results reproduce the observed gain compared to the baseline system over an additive white Gaussian noise (AWGN) channel without OFDM. Obviously, this provides an upper performance bound of the OFDM autoencoder.
2) *Explicit pilot without equalizer:* In this step we remove the MMSE equalizer block and leave the task of equalizing to the NN. However, the autoencoder is still provided with the pilot, i.e., the receiver input dimension (per subcarrier) is $\frac{n}{2} + 1$.
3) *No pilot and no equalizer:* Interestingly, the system shows a slight performance improvement with respect to the baseline although it does not require a dedicated pilot and operates at a different rate.[2]

This leads to the conclusion that the autoencoder does neither require explicit pilots nor explicit equalization, or, in other words, the autoencoder can efficiently learn to equalize for a single-tap channel.

### C. Piloting

As shown above, the autoencoder does not need any explicit piloting. This raises, however, the question how the autoencoder handles equalization. Therefore, in Fig. 6, we compare the constellation diagrams of an autoencoder trained with explicit pilot and the same setup without explicit piloting. Interestingly, the autoencoder is able to learn some kind of *superimposed-piloting* over all symbols by shifting the center of the constellations.

---
[2]We neglect the rate loss due to piloting in the SNR calculation as the fraction of pilots is somehow arbitrary. However, when considering this rate loss, the autoencoder without explicit pilot performs comparable to the autoencoder with explicit pilot.

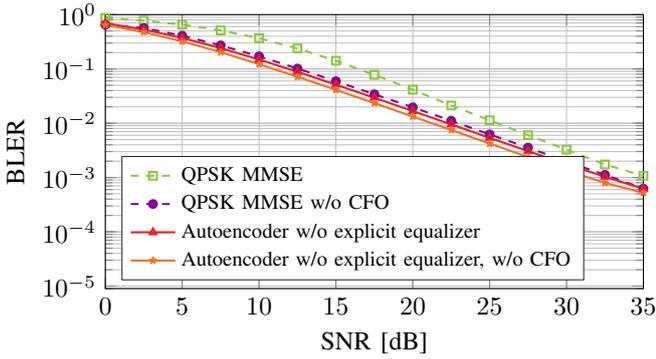

Fig. 7: Effect of CFO to the BLER performance of the autoencoder with RTN-based compensation and the QPSK baseline with conventional CFO compensation. The oscillator accuracy is assumed to be 20 ppm in both transmitter and receiver (i.e., in total 40 ppm maximum offset).

*D. CFO Compensation*

Next, we consider an CFO affected channel and use another RTN (not shown for space reasons) to estimate the phase offset of the complete signal in time-domain. For this experiment, both systems use an explicit pilot and the baseline uses a conventional CFO compensation in the time-domain. The results in Fig. 7 show that the autoencoder can adapt to the CFO with smaller performance degradation than the baseline.

*E. Effects of Non-Linearities*

To demonstrate the flexibility and simplicity of the autoencoder concept in accounting for new channel effects, we now introduce non-linear amplifiers and clipping. We model this effect as AM-AM distortion by a third-order non-linear function with normalized input as given in [11]

$$g(x) = x - \alpha |x|^2 x$$

with $x \in \mathbb{C}$ and $|x|$ in $[0, 1]$, i.e., we clip $x$ such that $|x| \leq 1$ and keep the phase of $x$ unchanged. We choose $\alpha = 1/3$ and scale the input $g(x)$ to realize a certain error vector magnitude (EVM). For the baseline, we assume an EVM of -10.97 dB, which is the LTE requirement for 64-QAM [12, Sec. 14.3.4].

As expected and depicted in Fig. 8, the baseline shows a degraded performance. However, when directly training the autoencoder on the new channel model, the performance improves significantly without the need for any further compensation algorithms. We want to emphasize that the significance of this experiment is not just the fact that the autoencoder can easily handle this non-linearity in the channel. The crucial point is that this approach allows considering and compensating multiple effects straightforwardly. This makes the concept of end-to-end learning a universal and flexible tool for future system design.

## V. CONCLUSIONS AND OUTLOOK

We have shown that the autoencoder system can be embedded into an OFDM with CP system. This mitigates synchronization issues and also simplifies equalization over multipath channels, leading to practical communication systems with feasible training and inferencing complexity. Although being difficult to analyze, we have shown that the system *learns* to use pilots if required. Furthermore, we observe that equalization is done equally well as in a conventional MMSE equalizer. Also CFO can be handled directly in the time domain by an additional RTN. The benefits of such a system may be a reduced complexity (keeping in mind that the whole receiver is essentially described by a few matrix multiplications and additional non-linear activation functions) and more flexibility regarding imprecise knowledge about the channel. Finally, we have shown that hardware imperfections such as non-linearities can be easily considered into the whole system design and effectively compensated without larger efforts, further emphasizing the conceptual simplicity of end-to-end learning.

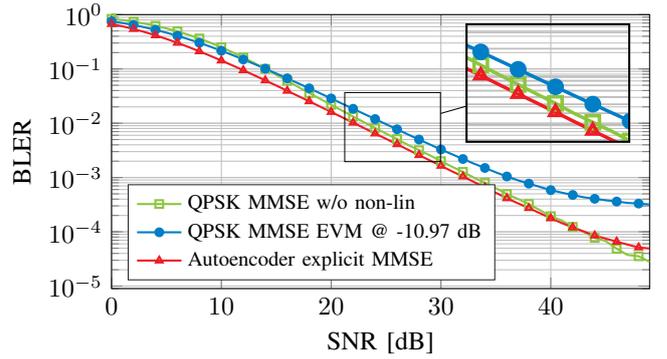

Fig. 8: Impact of non-linear amplifiers ($\alpha = 1/3$) in combination with clipping. The baseline operates at an EVM of -10.97 dB.